# Dendritic voltage recordings explain paradoxical synaptic plasticity: a modeling study

Abbreviated title: Dendritic Voltage-based Plasticity Rule

Claire Meissner-Bernard* (1), Matthias Chinyen Tsai (2), Laureline Logiaco (3), Wulfram Gerstner (2)
 (1) Friedrich Miescher Institute for biomedical research, 4058 Basel, Switzerland
 (2) EPFL, 1015 Lausanne, Switzerland
 (3) Columbia University, Center for Theoretical Neuroscience, 10027 New York, USA

*Materials & Correspondence: claire.meissner-bernard@fmi.ch

## Abstract

Experiments have shown that the same stimulation pattern that causes Long-Term Potentiation in proximal synapses, will induce Long-Term Depression in distal ones. In order to understand these, and other, surprising observations we use a phenomenological model of Hebbian plasticity at the location of the synapse. Our model describes the Hebbian condition of joint activity of pre- and postsynaptic neurons in a compact form as the interaction of the glutamate trace left by a presynaptic spike with the time course of the postsynaptic voltage. Instead of simulating the voltage, we test the model using experimentally recorded dendritic voltage traces in hippocampus and neocortex. We find that the time course of the voltage in the neighborhood of a stimulated synapse is a reliable predictor of whether a stimulated synapse undergoes potentiation, depression, or no change. Our computational model can explain the existence of different -at first glance seemingly paradoxical- outcomes of synaptic potentiation and depression experiments depending on the dendritic location of the synapse and the frequency or timing of the stimulation.

# 1. Introduction

How are memories encoded in the brain? In 1949, Donald Hebb postulated that a synapse connecting two neurons strengthens if both neurons are active together (Hebb, 1949). Numerous experiments have confirmed the interaction of pre- and postsynaptic neuronal activity during the induction of synaptic plasticity (see Levy & Stewart, 1983; Bliss & Collingridge, 1992; Sjöström et al., 2001; Wang et al., 2005; Markram et al. 2011). The critical postsynaptic signal for plasticity induction might be related to voltage (Artola et al., 1990), calcium (Cormier et al., 2001), or backpropagating action potentials (Markram et al., 1997b). If changes in subthreshold voltage or calcium concentration are the critical signals on the postsynaptic side, then plasticity does not require the postsynaptic neuron to fire a somatic spike. On the other hand, if backpropagating action potentials are critical, then synaptic plasticity outcomes can be completely described by the relative timing of pre- and postsynaptic spikes in the form of a Spike-Timing Dependent Plasticity rule (STDP rule, Abbott and Nelson, 2000).

A first, and fundamental, challenge for all STDP models is the existence of subthreshold plasticity in the absence of somatic spikes (Ngezahayo et al., 2000; Golding et al., 2002; Brandalise et al., 2014; Lisman & Spruston, 2005). Another challenge for some (Gerstner et al., 1996a; Song et al., 2000; Kistler & van Hemmen, 2000), but not all (Senn et al., 2001; Pfister & Gerstner, 2006; Clopath & Gerstner, 2010; Graupner & Brunel, 2012) STDP models is the interaction of frequency and spike-timing dependence so that long-term potentiation (LTP) for pre-before-post timing disappears at low frequencies (Sjöström et al., 2001; Markram et al., 1997b) whereas long-term-depression (LTD) for post-before-pre timing does not (Sjöström et al., 2001). Finally, an important finding that challenges classical models of STDP (Gerstner et al., 1996a; Song et al., 2000; Van Rossum et al., 2000) is the observation that plasticity rules depend on synapse location: whereas normally a protocol of presynaptic stimulation followed by postsynaptic activity induces potentiation, it was found to induce depression in distal synapses (Froemke et al., 2005; Letzkus et al., 2006). We refer to the above challenges as paradoxical effects of STDP and ask whether a single phenomenological model can account for all of these.

The observations that plasticity depends on dendritic synapse location and does not require somatic spikes hint at dendritic effects that are not accounted for by standard STDP models. Indeed, dendritic spikes have been shown to play a key role for the induction of plasticity in various brain regions (Holthoff et al., 2004; Kampa et al., 2006 and 2007; Gambino et al., 2014; Remy & Spruston, 2007). Dendritic events are linked to active channel properties which can vary along the dendritic tree (Spruston, 2008). Local dendritic nonlinearities could thus explain why different learning rules can be obtained with similar protocols in different brain regions or even within the same cell as a function of synapse location.

However, it is difficult to translate such an insight into a concrete biophysical model because it would require as a starting point a valid, and broadly accepted, model of local dendritic nonlinearities as well as a biophysically plausible synaptic plasticity model – but neither of these are readily available. While a first step in this direction has been taken recently (Ebner et al. 2019), most of the biophysical and phenomenological plasticity rules proposed over the years have in practice been tested using simplified point neuron models. And even if a biophysically detailed nonlinear dendrite model with active zones (Hay et al. 2011) were to be used, the location, and composition of ion channels in such active zones, might not be exactly the one encountered in the specific neuron recorded in an experiment.

In this paper we propose a voltage-based plasticity model that can be fitted to experiments without the need of fine-tuning any biophysical neuron model. Our model can be seen as a variation of earlier phenomenological voltage-based (Brader et al., 2007, Clopath et al., 2010) and calcium-based plasticity models (Shouval et al., 2002; Graupner & Brunel, 2012; Rubin et al., 2005). Our model has a set of plasticity parameters that need to be tuned. However, tuning of additional neuronal parameters is not necessary simply because we do not use any biophysical neuron model but work directly with the experimentally measured time course of the voltage in the neighborhood of the synapse.

In this paper we investigate whether such a phenomenological model of synaptic plasticity in which the arrival of neurotransmitter is paired with the postsynaptic voltage at the location of the synapse can explain the aforementioned paradoxical experimental results on excitatory synapses: (i) LTP in the absence of somatic spikes; (ii) the interaction of spike-timing and spike frequency; and (iii) inversion of a plasticity rule as a function of dendritic location. We focus on three experiments where the time course of dendritic voltage was measured during the application of a plasticity-inducing protocol (in neocortex, Letzkus et al., 2006 and in hippocampus, Brandalise et al., 2014 and 2016). We show that the time-course of the postsynaptic voltage in the neighborhood of the synapse, in combination with presynaptic signaling, is a reliable predictor of synaptic plasticity and is sufficient to explain the outcome of the experiments. In brief, our voltage-based model replicates plasticity behaviors of synapses across various dendritic locations in neocortex and hippocampus.

## 2. Results

### Voltage dependence of plasticity

Our model combines ideas from phenomenological models of voltage-based plasticity (Brader et al., 2007, Clopath et al., 2010) with the 'veto' concept of Rubin et al. (2005). As described in the Methods section, each presynaptic spike leaves, in our model, a trace $\bar{x}$ at the synapse; analogously, the activity of the postsynaptic neuron also leaves two traces at the synapse, described as two low-pass filtered version $\bar{u}_+$ and $\bar{u}_-$ of the dendritic voltage u. Potentiation can occur if the variable $\bar{u}_+$ (i.e., the voltage filtered with time constant $\tau_+$) is above some threshold $\theta_+$. Similarly, depression can occur if $\bar{u}_-$ (i.e., the voltage filtered with time constant $\tau_-$) is above some threshold $\theta_-$. In both cases, the amount of change depends on the momentary value of the trace $\bar{x}$ left by a presynaptic spike (**Figure 1** A-C). Importantly, to translate the competition between the molecular actors involved in LTP and LTD (phosphatase *vs.* kinase, see Bhalla & Iyengar, 1999; Xia & Storm, 2005; Herring et al, 2016) into mathematical equations, we introduce into our model a 'veto' concept: a potentiation signal overwrites LTD that would occur otherwise (Rubin et al., 2005; O'Connor et al., 2005; Cho et al., 2001). In our model, the veto mechanism is implemented by a dynamic LTP-dependent increase of the LTD-threshold $\theta_-$ that is characterized by parameters $b_\theta$ and $\tau_\theta$.

If we pair presynaptic stimulations with a constant voltage at the location of the synapse, our model shows three regimes (**Figure 1** D-G): (i) for hyperpolarization or voltage close to rest, synapses do not show any plasticity; (ii) for voltages above a first threshold $\theta_0$, presynaptic stimulation leads to a depression of the synapses; (iii) for voltages above a second threshold $\theta_1$ the synapses exhibit potentiation. Depending on the parameters, the voltage-plasticity relationship can be linear (**Figure 1** E) or non-linear (**Figure 1** F). Our model is consistent with experimental results of Ngezahayo et al. (2000) who paired 2 Hz presynaptic stimulations with constant postsynaptic depolarizations (voltage clamp) and determined the stationary voltage dependence of LTP and LTD induction. For a wide range of parameter choices, our model qualitatively reproduces this voltage dependence, if we assume that during clamping the voltage u at the dendrite is equal to the somatic voltage.

In more realistic experiments, the voltage at the location of the synapse is not constant but changes as a function of time. In the following, we investigate if our model can reproduce the experimentally measured plasticity observed with various LTP- or LTD-inducing protocols. We focus on experimental paradigms where the dendritic voltage was recorded close to the stimulated synapse during plasticity induction (Letzkus et al., 2006, Brandalise et al., 2014 and 2016). In brief, the methodological approach is as follows: we feed dendritic voltage time courses and corresponding presynaptic spike trains into our plasticity model. The variable u(t) is thus the experimentally recorded voltage, and not a simulated one. Parameters of our model are then fitted so that the plasticity predicted by our model matches as closely as possible the experimental plasticity values. To this end, we use an optimization algorithm which minimizes the mean-squared error (squared difference between theoretical and experimental plasticity outcomes, see Methods).

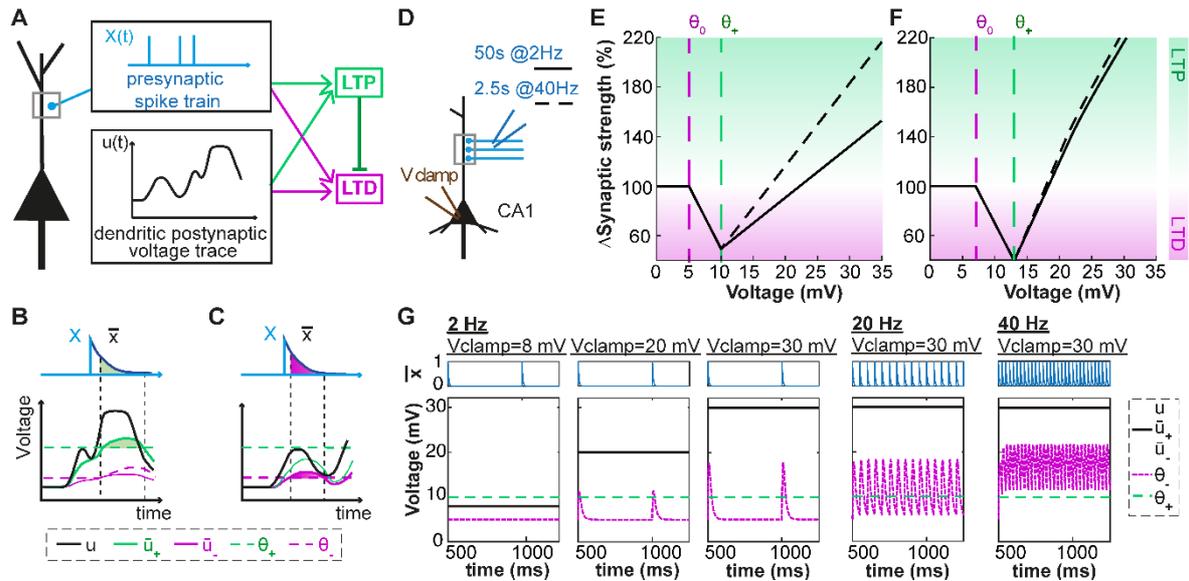

**Figure 1. Voltage-dependent plasticity model. (A)** The activity X(t) of the presynaptic neuron induces local dendritic voltage changes u(t) in the postsynaptic neuron. The change in synaptic strength (LTP or LTD induction) depends on the timing of the presynaptic spike, and the voltage close to the synapse. **(B)** The presynaptic spike X(t) leaves a trace $\bar{x}(t)$ at the synapse. The voltage u is low-pass filtered with a time constant $\tau_+$ (for the variable $\bar{u}_+$) or $\tau_-$ (for $\bar{u}_-$). The amount of LTP is proportional to $\bar{x}$ multiplied by $\bar{u}_+$, while $\bar{u}_+$ is above a threshold $\theta_+$. **(C)** Similarly, the amount of LTD is proportional to $\bar{x}$ multiplied by $\bar{u}_-$, while $\bar{u}_-$ is above a threshold $\theta_-$ which is lower than $\theta_+$ and increases when LTP occurs. **(D-G)** Plasticity in hippocampal model cells: extracellular afferent stimulation is paired with voltage-clamp of the postsynaptic neuron at different potentials (see Ngezahayo et al., 2000). 100 brief extracellular afferent stimulations are done at 2 different frequencies: 2 (full line) and 40 Hz (dotted line). **(E-F)** Synaptic strength w in percentage of its initial value as a function of voltage with respect to resting potential, in mV, for two sets of parameters. **(G)** The presynaptic trace $\bar{x}$ (blue) and the voltage u superimposed with its filtered versions $\bar{u}_+$ and $\bar{u}_-$ (black) for 3 different values of clamped voltage (8, 20 and 30 mV) and different stimulation frequencies. The thresholds are indicated by the dashed green ($\theta_+$) and dashed purple ($\theta_-$) lines. The parameters are: **(E and G)** $\tau_x$ =5ms, $\tau_+$ =6ms, $\tau_-$ =15ms, $\theta_+$=10mV, $\theta_0$=5mV, $A_{LTP}$=0.0001mV$^{-1}$.ms$^{-1}$, $A_{LTD}$=0.0001mV$^{-1}$.ms$^{-1}$, $b_\theta$=31000mV.ms, $\tau_\theta$= 14ms. **(F)** $\tau_x$ =5ms, $\tau_+$ =7ms, $\tau_-$ =15ms, $\theta_+$=13mV, $\theta_0$=7mV, $A_{LTP}$=0.0001mV$^{-1}$.ms$^{-1}$, $A_{LTD}$=0.0001mV$^{-1}$.ms$^{-1}$, $b_\theta$=45000mV.ms, $\tau_\theta$ =5ms.

## Subthreshold plasticity in the hippocampus

We first used experimental data from rat hippocampus where Brandalise et al., 2014 investigated plasticity at the CA3 recurrent synapses using a subthreshold protocol. In this protocol, an excitatory postsynaptic potential (EPSP) of a few millivolts induced by stimulation of the CA3 recurrent pathway was paired with a subthreshold mossy fiber (MF) stimulation that, if stimulated separately, also led to an EPSP of a few millivolts (Figure 2A). Dendritic voltage recordings (data kindly shared by F. Brandalise) were used as the variable u(t) in our plasticity model (see Material and Methods).

In agreement with the experimental results of Brandalise et al. (2014), no plasticity was induced in our model synapses when there was no MF stimulation (CA3 alone) or when the MF and CA3 stimulations occurred at the same time (0 ms). LTD was observed when the MF stimulation preceded the CA3 stimulation with a 40 ms interval (noted as -40 ms in **Figure 2** D,H). LTP was observed when the MF stimulation followed the CA3 stimulation with a 10 ms time interval (+10 ms).

Importantly, in the experiments of Brandalise et al. (2014), repetitive pairings at +10ms were found to induce two types of dendritic voltage trajectories (**Figure 2** B): either a nearly linear addition of the EPSPs caused by MF and CA3 stimulation or (in 31 ± 4 % of cases) a strongly supralinear voltage response (Brandalise et al., 2014). For example, a given cell could exhibit during one paired stimulation a linear response, and the same cell could then show in the next stimulation a supralinear response. Since our model uses the experimental voltage time course, we could predict the plasticity outcome for a given cell (**Figure 2** H) based on the dendritic voltage recorded for that specific cell across different conditions (CA3 alone, 0 ms and +10 ms protocols, see Material and Methods). Hence variations in the amount of plasticity are explained in the model by differences in the voltage recordings – without any tuning of parameters between different cells. In particular, when the occurrence of supralinear events was completely blocked, no plasticity was observed in our model synapse (10 ms, block, **Figure 2** H), in agreement with experiments (Brandalise et al., 2014).

To understand how the model works, let us focus on a few examples (**Figure 2** B-D). During the -40 ms protocol, the low-pass filtered voltage trace $\bar{u}_+$ did not reach the threshold $\theta_+$ for LTP induction, whereas the voltage $\bar{u}_-$ filtered with a larger time constant reached $\theta_-$, inducing LTD (**Figure 2**D). With the +10 ms protocol, $\bar{u}_+$ reached $\theta_+$ only during trials in which a supralinear event occurred. During linear events, $\bar{u}_+$ and $\bar{u}_-$ did not reach their respective thresholds $\theta_+$ and $\theta_-$ (**Figure 2**B). This was also the case during the 0 ms protocol (**Figure 2**C).

## STDP protocol in the hippocampus and cross-validation procedure

In a further set of experiments using a burst STDP protocol, Brandalise et al. (2016) paired a recurrent CA3 EPSP with 3 action potentials (APs) at 200 Hz (10 ms time interval, see **Figure 2** E). This stimulation led in more than half of the trials to the generation of a dendritic spike (**Figure 2** F), unless a hyperpolarizing step current was applied in the dendrite during the brief somatic injections triggering the APs. Similarly, pairing the CA3 EPSP with 3 APs at 50 Hz or with a single AP did not generate a dendritic spike (**Figure 2** G). Representative dendritic voltage traces u(t) measured by Brandalise et al. (2016) were used in our plasticity model which was able to reproduce the plasticity outcomes of the experiments (**Figure 2** H).

We emphasize that our plasticity model with a *fixed set* of parameters (**Table 1**) could reproduce the outcome of all the STDP experiments as well as that of all the earlier subthreshold protocols (**Figure 2** H). The set of parameters in **Table 1** was obtained with all available voltage traces corresponding to 15 plasticity outcomes.

|  | $\tau_x$ [ms] | $\tau_+$ [ms] | $\theta_+$ [mV] | $\theta_0$ [mV] | $A_{LTP}$ [mV$^{-1}$.ms$^{-1}$] | $A_{LTD}$ [mV$^{-1}$.ms$^{-1}$] | $\tau_-$ [ms] | $b_\theta$ [mV.ms] | $\tau_\theta$ [ms] | LSE |
|---|---|---|---|---|---|---|---|---|---|---|
| Letzkus (Fig.3) | 22.4 | 2.00 | 27.1 | 6.20 | 4.27x10$^{-5}$ | 16.5x10$^{-5}$ | 60.0 | 1.00x10$^4$ | 29.1 | 7.2x10$^{-2}$ |
| Brandalise (Fig.2) | 14.3 | 7.80 | 9.94 | 4.04 | 225x10$^{-5}$ | 691x10$^{-5}$ | 53.3 | 9.91x10$^{-1}$ | 1.99 | 9.3x10$^{-3}$ |
| Sjostrom (Fig.4) | 5.08 | 17.8 | 11.8 | 6.50 | 37.2x10$^{-5}$ | 31.2x10$^{-5}$ | 24.9 | 24.7x10$^4$ | 2.49 | 2.6x10$^{-1}$ |

**Table 1. Parameters minimizing the error (see Methods).** The least-square error (LSE) is defined as the squared difference between the experimental and theoretical plasticity values summed over various protocols.

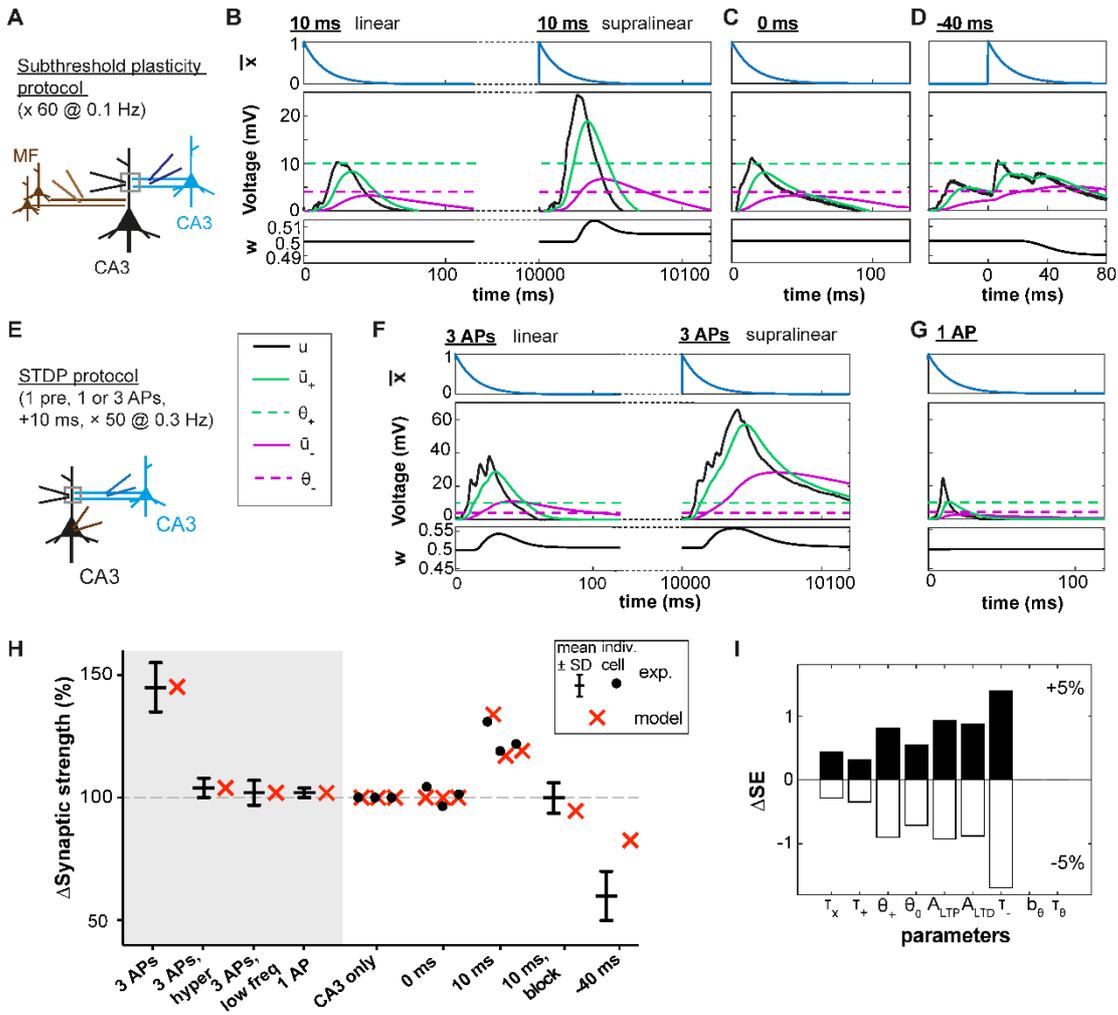

**Figure 2. Subthreshold and spike-timing dependent plasticity at CA3 synapses in the hippocampus. (A)** Experimental setup. Stimulation of CA3 recurrent inputs (blue electrode) was paired with a subthreshold stimulation of mossy fiber inputs (MF, brown electrode). The pairing is repeated 60 times at 0.1Hz (Brandalise et al., 2014). **(B-D)** Experimental voltage traces (black, middle panel) caused by subthreshold stimulations with a 10ms **(B)**, 0 ms **(C)**, or -40 ms **(D)** interval. Blue traces (top panels, $\bar{x}(t)$), green and purple (middle panels: $\bar{u}_+$ green full line; $\bar{u}_-$ purple full line; $\theta_+$ green dashed line; and $\theta_-$ purple dashed line), and black (bottom panels, w) show the time course of selected model variables during the simulated experiments. Two different cases are illustrated in **(B)**, because for the same stimulation two types of voltage responses were recorded in the dendrite (black electrode in A): linear (left) and supralinear (right) ones. The supralinear responses correspond to the occurrence of dendritic spikes. **(E)** STDP protocol (Brandalise et al, 2016): stimulations of CA3 recurrent inputs were paired 50 times with brief somatic current injections (2 ms; 4 nA) which evoked action potentials (APs). **(F)** When 3 APs were evoked at a frequency of 200 Hz, dendritic spikes (black, right panels) occurred in 60% of the trials. In the remaining 40%, a linear response was generated (left); color of other traces as in B and C. **(G)** Responses were always linear, if only one AP was paired with stimulation of CA3 recurrent inputs. **(B, C, D, F, G)** Time of presynaptic stimulation is set to 0. **(H)** Plasticity outcome using different plasticity protocols. STDP protocol (see E): presynaptic stimulation paired either with 3 APs (see F), with 3 APs and concomitant application of a hyperpolarizing pulse, with 3 APs generated at a lower frequency (50 Hz) or with 1 AP (see G). Subthreshold protocol: CA3 stimulation only, no pairing; pairings with different time intervals (0 ms, 10 ms, -40 ms, see B-D) and, for the 10 ms time interval, simultaneously blocking the occurrence of supralinear events (10ms, block). Filled circles represent data from individual cells. Black error bars represent experimental mean +/- SD. Red crosses represent simulations using the parameters obtained with the best fit. All data points in H are fitted with a single set of parameters (see text and Table 1). Differences in plasticity for the same protocol (e.g. +10 ms) arise due to differences in experimental voltage traces. Data shared by F. Brandalise. **(I)** Squared Error (SE) of the best fit subtracted from the SE obtained after increasing (upper panel) or decreasing (lower panel) each parameter by 5%, one parameter at a time (ΔSE).

Since our model has 9 free parameters, the question arises whether the model is overfitting the available data points or whether it would correctly generalize to novel data. In order to check the model's predictive power, we used an additional, independent, optimization procedure (leave-one-out cross-validation): we fitted the model parameters on plasticity outcomes for 14 voltage traces by minimizing the mean-squared error and predicted the plasticity outcome on the remaining trace (see Table 2 for the statistics over all 15 leave-one-out experiments). Even though the median error after testing the plasticity outcome on the excluded voltage traces was (as expected) larger than the median training error (Table 2), its actual value of $1.5 * 10^{-3}$ was comparable to the normalized error of $6.2 * 10^{-4}$ observed in the direct fitting approach of **Table 1**. Furthermore, we found that most parameter values are consistent across the 15 leave-one-out experiments as indicated by a small standard deviation of the parameter value compared to its mean value (Table 2); exceptions were the veto parameters $b_\theta$ and $\tau_\theta$ which showed rather large standard deviations. A sensitivity analysis further confirmed that the exact values of these two parameters was not critical (Figure 2I and Supplementary Figure 1). Thus, cross-validation and sensitivity analysis confirm that the model has predictive power.

| LSE | | Coefficient of variation (%) | | | | | | | | |
|---|---|---|---|---|---|---|---|---|---|---|
| Training (normalized) | Testing | $\tau_x$ | $\tau_+$ | $\theta_+$ | $\theta_0$ | $A_{LTP}$ | $A_{LTD}$ | $\tau_-$ | $b_\theta$ | $\tau_\theta$ |
| $6.3 \times 10^{-4}$ | $1.5 \times 10^{-3}$ | 8.0 | 5.9 | 3.4 | 13 | 13 | 22 | 14 | 130 | 110 |

**Table 2. Cross-validation results.** First two columns: Median error of the model after training (1st column) on 14 plasticity traces of the Brandalise experiments and testing (2nd column) on the 15th excluded one. Remaining columns: coefficient of variation for each parameter (sd/mean*100) across the 15 sets of best parameters found during the cross-validation procedure.

## Location-dependent plasticity in neocortical apical dendrites

We next tested our model on data recorded at synapses between layer 2/3 and layer 5 pyramidal neurons in slices from rat somatosensory cortex (Letzkus et al., 2006). We fed our plasticity model with representative dendritic voltage traces recorded close to synapses located 100μm (proximal), 330μm or 660μm (distal) away from the soma (Figure 3A-C, Letzkus et al., 2006). Using the fitting procedure described above, we found that our voltage-based plasticity model with a single set of parameters could account for the plasticity results obtained with a burst STDP protocol (see **Figure 3** D and **Table 1**). Consistent with experimental results, the observed plasticity results varied in the model depending on synapse location on the dendritic tree. Proximal EPSPs were potentiated during pairings with somatic bursts of APs occurring 10 ms after the onset of the EPSP (+10 ms) and depressed when the postsynaptic bursts occured 10 ms before the EPSP (-10ms); at distal synapses, however, the pattern was reversed and EPSPs depressed during +10 ms pairings and were potentiated for -10 ms pairings (Letzkus et al., 2006).

Moreover, the model with the same set of parameters could also explain why distal EPSPs no longer potentiated after pairings at -10ms but still depressed during pairings at +10 ms, if the amplitude of the dendritic spikes evoked by the AP bursts decreased due to the presence of $NiCl_2$ (a blocker of a subtype of voltage-gated calcium channels, **Figure 3** C and E, and Letzkus et al., 2006).

To understand the workings of our model, we observed different model variables as a function of time. At distal synapses, during +10 ms pairings, the value of the presynaptic trace x̄ had already decreased significantly when ū₊ reached the threshold θ₊ (**Figure 3** A). The amount of LTP was therefore not high enough for the veto to have a significant impact on LTD induction. As a result, weak LTD occurs (**Figure 3** D). In contrast, for -10 ms pairings, x̄ switched from 0 to its maximal value 1 at a moment when $\bar{u}_+$ was close to its maximal value well above $\theta_+$ (**Figure 3** B). Therefore, the amount of LTP induced was high. The large LTP signal vetoed the induction of LTD as manifested by an increase in the LTD threshold $\theta_-$. As a result, LTP dominates, in agreement with experiments (**Figure 3** D). However, for -10 ms pairings, in the presence of $NiCl_2$ or at proximal synapses, the difference between $\bar{u}_+$ and $\theta_+$ was significantly reduced compared to what was observed at distal synapses, leading to an absence of synaptic potentiation (**Figure 3** B-C). Moreover, blocking of the veto-mechanism reduces the quality of the fit (**Table 3**).

Thus, the results of our voltage-based plasticity model support the idea that differences in the voltage traces can explain the spatial differences in the learning rule, as suggested by Letzkus et al. (2006). Importantly, all the above experiments (Letzkus et al. 2006) are explained by the same model with the same set of parameters (**Table 1**).

|  | Error (with veto) | Error (without veto) | Error (with veto)/Error (without veto) |
|---|---|---|---|
| Letzkus | $7.2 \times 10^{-2}$ | $12 \times 10^{-2}$ | 0.60 |
| Brandalise | $9.3 \times 10^{-3}$ | $9.3 \times 10^{-3}$ | 1.0 |
| Sjostrom | $2.6 \times 10^{-1}$ | $3.6 \times 10^{-1}$ | 0.72 |

**Table 3. Parameters minimizing the error with and without the veto term.** The same optimization procedure was run without the veto terms $b_\theta$ and $\tau_\theta$.

## High-frequency pairings in neocortical basal dendrites

We have until now focused on plasticity results obtained after repeated pairings of pre and postsynaptic activities at a low frequency (0.1 and 1 Hz). Yet, an important feature of synaptic plasticity is its frequency-dependence. Different amounts of plasticity are obtained by repeating the same pairings at different frequencies. Unfortunately, experimental dendritic recordings do not exist for these types of experiments. Results have been obtained among others at L5-L5 synapses of rat neocortical neurons (Sjöström et al., 2001), which are well-characterized. More than half of the synaptic contacts between L5 neurons are made on basal dendrites ($80 \pm 35$ μm from the soma in young rats, Markram et al., 1997a), for which a detailed model exists (Nevian et al., 2007). We simulated dendritic voltage around 80 μm from the soma using the aforementioned neuron model at four different frequencies (0.1, 10, 20 Hz and 40 Hz) and with different time intervals between the presynaptic and the postsynaptic stimulation (-10, 0, +10 and +25 ms). As shown in **Figure 4**, our model with fixed set of parameters can reproduce both the frequency-dependence and spike-timing dependence of plasticity.

At low frequencies, the time between two pairings is long enough so that the membrane potential u repolarizes back to its resting value. As a consequence, $\bar{u}_+$ is close to zero when the next pairing occurs. This is not the case at high frequencies: the residual depolarization between postsynaptic spikes allows $\bar{u}_+$ to reach the threshold $\theta_+$, leading to LTP induction (see **Figure 4** D). Similarly, a correlation between the amount of residual depolarization and the amount of LTP has been found in Sjostrom et al. (2001). Interestingly, the spike after-depolarization of the neuron model by Nevian et al. (2007) seems to have a shorter time constant than the one recorded in Sjostrom et al. (2001, see **Figure 4** B). This explains why in our model no LTP is induced for pairing frequencies below 20Hz. At high frequencies, the veto mechanism tunes down LTD (see **Figure 4** D). Blocking of the veto-mechanism reduces the quality of the fit (**Table 3**).

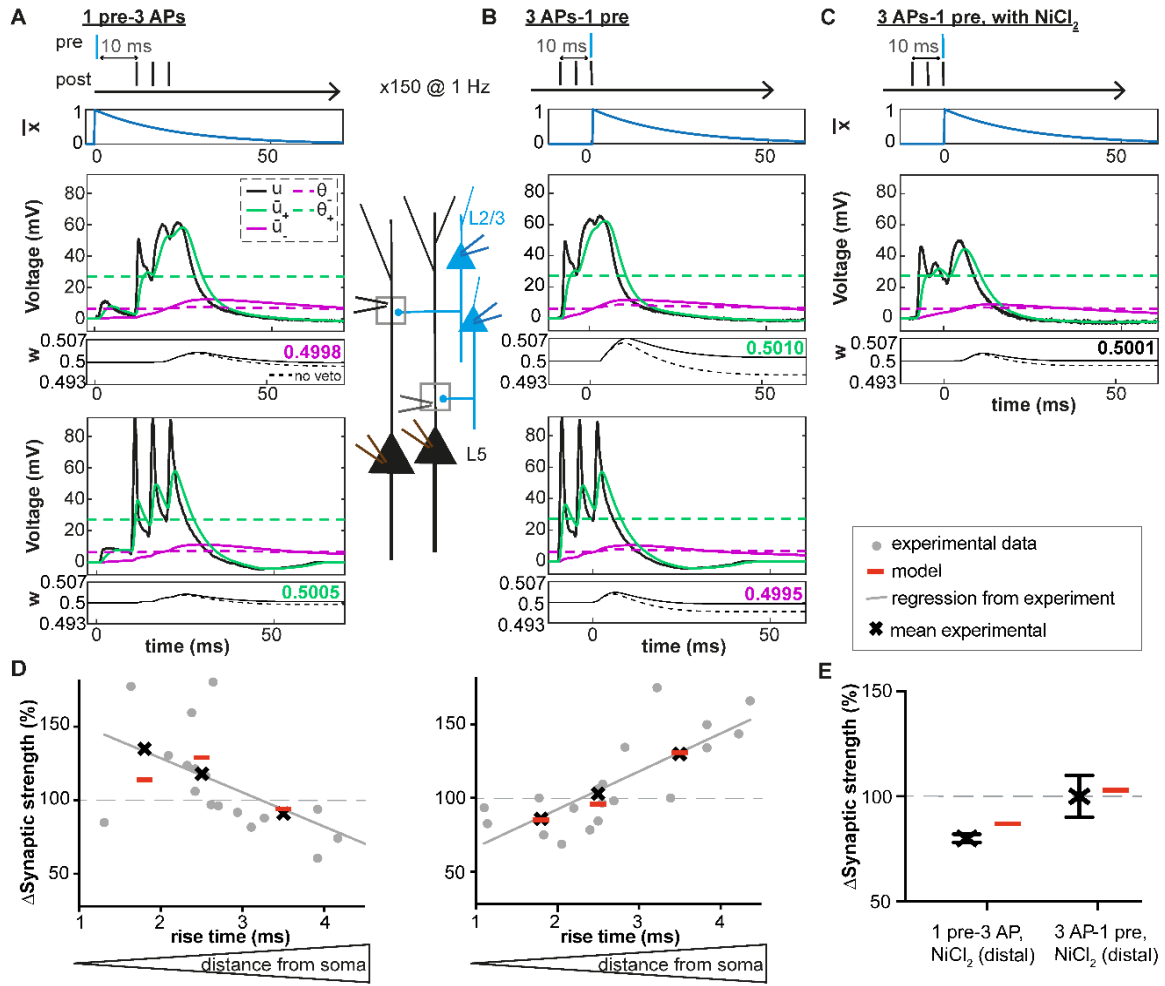

**Figure 3. Distance-dependent STDP at synapses between layer 2/3 and layer 5 pyramidal neurons in somatosensory cortex. (A-C)** Voltage traces at distal (middle) and proximal (bottom) synapses. Postsynaptic bursts (3 action potentials, APs at 200 Hz) are paired with presynaptic action potentials (±10 ms time interval, pairing frequency of 1 Hz, 150 repetitions). Experimental voltage traces u (black line) are redrawn from Letzkus et al. (2006). At distal synapses, dendritic spikes are generated. Glutamate trace $\bar{x}$ (blue line, top), filtered versions of the voltage $\bar{u}_+$ and $\bar{u}_-$ (solid green and purple lines), and synaptic weight w (number: value of w after one pairing; dashed line: $b_\theta$ is set to 0, no veto) as a function of time. **(A)** During +10 ms pairings, the value of the presynaptic trace $\bar{x}$ has already decreased significantly before $\bar{u}_+$ reaches $\theta_+$. The amount of LTP is not high enough for the veto to have a significant impact on LTD induction. **(B)** In contrast, for -10 ms pairings, $\bar{x}$ switches from 0 to its maximal value 1 at around the time when $\bar{u}_+$ reaches its maximal value far above $\theta_+$ and $\bar{u}_-$ is slightly above $\theta_-$. Therefore, the amount of LTP induced is high and significantly reduces LTD via an increase of the LTD threshold $\theta_-$. **(C)** However, in the presence of $NiCl_2$ (blocks a subtype of voltage-gated calcium channels), the difference between $\bar{u}_+$ and $\theta_+$ is significantly reduced compared to the control case in B, leading to no change in the synaptic strength. **(D-E)** Experimental and predicted change in synaptic weight. Crosses and dots or crosses with error bars represent plasticity from Letzkus et al. (2006) and red lines simulations. **(D)** Plasticity along the dendrite for the protocols described in A-C: 1pre-3 APs (left) or 3 APs-1 pre (right). EPSP rise time at the soma is a proxy of the distance between the plastic synapse and the soma (a distance of 110 μm, 330 μm and 660 μm correspond to a rise time of 1.8 ms, 2.5 ms and 3.5 ms, respectively, see Letzkus et al., 2006). See text for more details. **(E)** Plasticity at distal synapses in the presence of $NiCl_2$.

Since we have a validated model of the L5 basal dendrites, we can predict the plasticity outcome for plasticity protocols with triplets of spikes at L5-L5 synapses. During triplet experiments, a presynaptic [postsynaptic] spike is triggered between the occurrence of two postsynaptic [presynaptic] spikes. As demonstrated experimentally by Wang et al. (2005), the outcome of triplet experiments is not a linear sum of the outcome of each pair of pre- and post-synaptic spikes taken separately. Among others, while a post-pre pair with a 5 ms interval can trigger LTD, addition of a postsynaptic spike 5 ms after the presynaptic spike leads to LTP (see **Table 4**). Note that the data of Wang et al. (2005) is from hippocampal neurons and that our model predicts that plasticity behaves somewhat differently in basal dendrites of L5 neurons compared to hippocampus.

To summarize, the same voltage-based plasticity model can account for three different series of experiments corresponding to four publications (Sjöström et al., 2001; Letzkus et al., 2006; Brandalise et al., 2014 and 2016). Importantly, the model parameters are slightly different for different synapse types, but each series of experiments from one synapse type is explained by a single set of model parameters (**Table 1**). In other words, model parameters are kept fixed across all experimental results in a given experimental preparation.

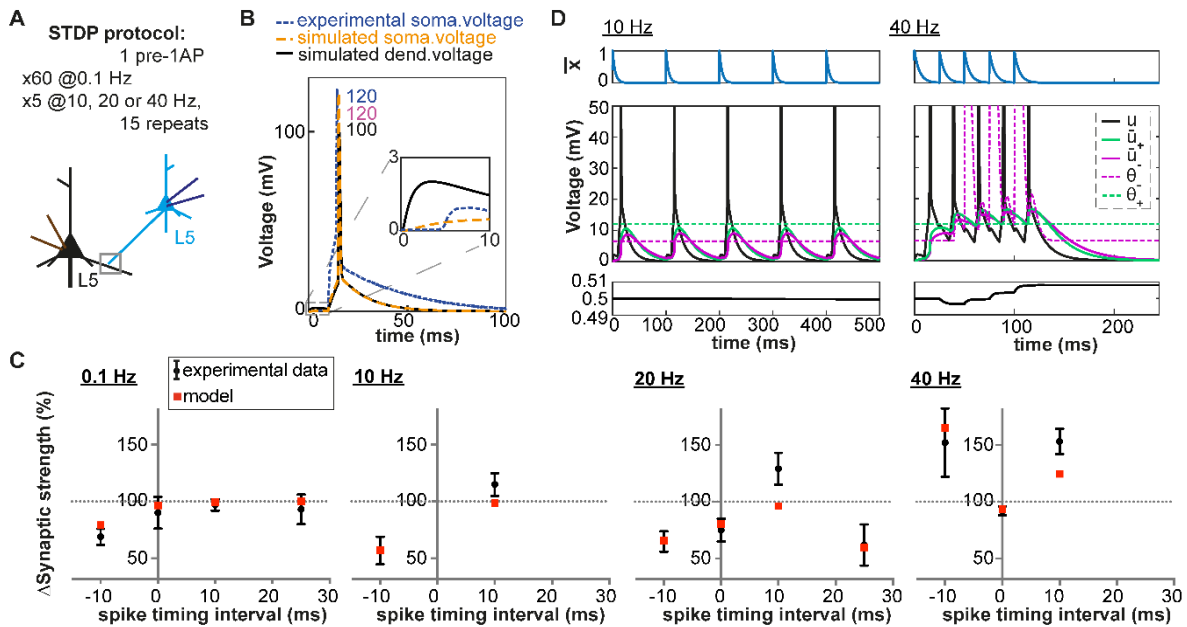

**Figure 4. Pairing and timing-dependence of plasticity at neocortical synapses. (A)** Two synaptically connected L5 neurons were stimulated with different time intervals (-10, 0, 10 and 25 ms) at different pairing repetition frequencies: 0.1 Hz, 10 Hz, 20 Hz and 40 Hz. **(B)** Simulated dendritic (black), somatic (orange) and experimentally recorded somatic (blue) voltage time course for +10 ms time interval (number: peak value). The experimental voltage trace is redrawn from Sjöström et al. (2001); inset: EPSP time course. **(C)** Plasticity as a function of spike timing. Each panel represents one pairing repetition frequency. LTP is induced at high frequencies. Black error bars represent data from Sjöström et al. (2001) and red squares represent our plasticity model. **(D)** Presynaptic trace $\bar{x}$ (blue), voltage $u$ (black) and its filtered versions $\bar{u}_+$ (green) and $\bar{u}_-$ (purple) for +10 ms time interval: 10 Hz **(left)** or 40 Hz **(right)** repetition frequency.

| Protocol | Spike timing interval (ms) | Cultured hippocampal cells, Wang et al. (2005) | Prediction for L5-L5 basal dendrites |
|---|---|---|---|
| Pre-post | 5 and 10 | LTP | No plasticity (98 and 99%) |
| Post-pre | 5 and 10 | LTD | LTD (89 and 78%) |
| Pre-post-pre | 5 | No plasticity | LTD (81%) |
|  | 10 | No plasticity | LTD (62%) |
| Post-pre-post | 5 | LTP | LTP (113%) |
|  | 10 | LTP | LTD (79%) |

**Table 4. Predicted plasticity for pairs and triplets of spikes**. The timing between pre and postsynaptic spikes is either 5 ms or 10 ms (60 repetitions at 1 Hz). Voltage traces were simulated using the Nevian et al. (2007) model and the plasticity model was simulated with the L5-L5 parameters. The results obtained in cultured hippocampal neurons (Wang et al., 2005) are reported as a comparison. Percentages indicate synaptic strength after application of the induction protocol compared to baseline.

## 3. Discussion

Long-term potentiation or long-term depression are induced through the combined action of the presynaptic and postsynaptic activities. We showed that a single phenomenological voltage-based model could explain results using various synaptic plasticity protocols: experiments (i) with voltage clamp (Figure 1); (ii) with variable time interval between presynaptic and postsynaptic spikes (Figures 2,3,4); (iii) with variable pairing frequency (Figure 4); (iv) with multiple postsynaptic spikes (Figures 2,3,4); (v) with subthreshold plasticity (Figure 2) and (vi) with location-dependence (Figure 3).

### Comparison with other plasticity models

The model proposed here, as well as other voltage-based and calcium based models (Shouval et al., 2002; Clopath & Gerstner, 2010; Graupner & Brunel, 2012, Brader et al., 2007, Rubin et al., 2005, Abarbanel et al., 2002), is a phenomenological one since it does not aim to describe the full mechanistic signaling chain from presynaptic spike arrival to a change in the number of AMPA receptors or presynaptic release probabilities. Rather it should be considered as a 'black-box' model that summarizes a large range of experimental results in the form of a compressed 'learning rule' with only a few variables.

In calcium-based models (Shouval et al., 2002; Rubin et al., 2005; Graupner & Brunel, 2012), calcium concentration acts as a summary variable that includes effects of both pre- and postsynaptic activity (influx of calcium through NMDA channels and other voltage-gated ion channels e.g.). The level (Shouval et al., 2002) or time course (Rubin et al., 2005, Graupner & Brunel, 2012) of the calcium concentration in the simulated model is then compared with threshold variables in order to predict occurrence of LTP or LTD. Rubin et al., (2005) proposed to add a "veto" of LTP on LTD when a relatively high calcium threshold was reached which inspired the veto mechanism in the present model.

In voltage-based models (Brader et al., 2007; Clopath & Gerstner, 2010), and similarly in our model, presynaptic activity leaves a filtered trace at the synapse. This trace can be interpreted for excitatory synapses as the amount of glutamate bound to postsynaptic receptors which induces calcium influx into the neuron (and/or, in the case of LTD, conformational changes of the receptors initiating intracellular signaling). It is this glutamate trace that interacts either with the postsynaptic voltage directly or with a low-pass filtered version thereof. The comparison of the voltage variables with several thresholds allows to predict the induction of LTP or LTD of those synapses that have been presynaptically stimulated (Brader et al., 2007; Clopath & Gerstner, 2010). Thus voltage-based models jump over the biophysics of calcium dynamics and connect the presynaptic stimulation in combination with the time course of the postsynaptic voltage directly with the outcome of plasticity experiments. We note that LTD could have a different mechanism (e.g., non-ionotropic, Nabavi et al., 2013) compared to LTP (ionotropic), while still consistent with our phenomenological model.

Our model and the model of Clopath et al. (2010) are different in some respects.
First, in Clopath et al., (2010), LTD is triggered by the joint action of an instantaneous spike event, rather than the glutamate trace and postsynaptic voltage. Using a glutamate trace (as opposed to a presynaptic spike event that covers a much shorter moment in time) is in our hands the only way to make LTD possible for pre-before-post pairings at low pairing frequencies (see Letzkus et al., 2006 as an example). Also, an extended glutamate trace looks biologically more plausible than a "point-like event" assumed in some classic STDP models (Song et al., 2000; Kistler & van Hemmen, 2000).
Furthermore, in Clopath et al., (2010), LTP can occur only if two conditions are met: the momentary voltage u(t) and the low-pass filtered voltage $\bar{u}_+$ need to be above a threshold $\theta_+$ and $\theta_-$, respectively. In other words, the membrane must already be depolarized before a spike occurs (see Sjöström et al., 2001). The two conditions together imply a quadratic dependence on voltage in the LTP inducing term (Clopath & Gerstner, 2010). Instead of a quadratic voltage term for LTP induction, our model works with a linear dependence on the thresholded, low-pass filtered voltage in combination with a veto-mechanism similar to the one suggested by Rubin et al. (2005).

Previous models were able to quantitatively fit the frequency dependence of STDP (Senn et al., 2001) as well as triplet and quadruplet effects of STDP protocols (Pfister & Gerstner, 2006; Clopath & Gerstner, 2010; Graupner & Brunel, 2012; Cai et al., 2006). The model of Graupner and Brunel (2012) also indicated how changes of STDP rules as a function of synaptic location on the dendrite could be qualitatively accounted for by changes of model parameters; in the absence of dendritic recordings and an appropriate dendrite model, a quantitative fit was not to be expected. A recent study by Ebner et al. (2019) used a detailed multicompartmental model of a neocortical neuron to simulate postsynaptic voltage at different dendritic locations and combined this voltage with a phenomenological model for four pathways of LTP and LTD induction. However, our voltage-based model is probably the first one to directly link dendritic voltage recordings with plasticity outcome, bypassing the need for a biophysically correct dendrite model.

In order to stabilize plasticity, the papers of Clopath et al. (2010) and Pfister and Gerstner (2006) suggest a form of metaplasticity implemented by a slow adjustment of the coefficients of LTD driven by the average mean firing rate. For network simulations we suggest to either implement an analogous slow adjustment of our LTD term as in Clopath et al. (2010) or keep parameters fixed and replace metaplasticity by heterosynaptic plasticity (Zenke et al., 2015 and 2017). The Clopath model also uses hard bounds for the weights which is recommended for network simulations.

## Role of dendritic spikes

Letzkus et al. (2006) showed that at distal locations, the peak amplitude of isolated backpropagating action potentials was half the size than that at proximal locations. Furthermore, postsynaptic bursts at the soma generated dendritic calcium spikes at distal locations. The two observations suggest that the somatic spike is less important for plasticity in distal dendrites than localized depolarizations at the location of the synapse. Similarly, in the hippocampal experiments of Brandalise et al. (2014 and 2016), dendritic NMDA spikes were generated: they resulted from high frequency bursting during the STDP protocol, and from broad and long mossy-fiber evoked EPSPs during the subthreshold protocol.

Both Letzkus et al. (2006) and Brandalise et al. (2014 and 2016) showed that LTP was abolished when dendritic spikes were blocked (pharmacologically or by hyperpolarizing the cell). In our model, the voltage time course at the location of the synapse determines whether or not LTP (or LTD) is induced at stimulated synapses. If the low-pass filtered voltage $\bar{u}_+$ does not reach a threshold $\theta_+$, then potentiation is impossible. Intrinsic dendritic nonlinearities of the postsynaptic neuron can boost voltage and explain the existence of different – at a first glance seemingly paradoxical - outcomes of plasticity experiments (Mishra et al., 2016). Since we paste the experimentally measured voltage traces directly into our plasticity model, the biophysical source of the depolarization does not matter.

## Predictions

Since our model is a phenomenological one (as opposed to a biophysical model that attempts to describe the full signal induction chain, e.g., Bhalla & Iyengar, 1999; Lisman & Zhabotinsky, 2001; Badoual et al., 2006; Saudargiene & Graham, 2015), it cannot be used as a predictive tool in cases where specific biochemical molecules are manipulated without affecting the voltage time course. However, one interesting qualitative prediction follows from the interaction of the veto-concept in our voltage based model. We predict a voltage-dependence of LTP induction (**Figure 1** E-F) that depends on the stimulation frequency of glutamate pulses. Since presynaptic vesicles are likely to deplete rapidly, we propose an experiment where presynaptic spike arrivals are replaced by glutamate puffs of standardized size while the postsynaptic voltage is clamped at a constant voltage. The prediction from our simple voltage-based model is that the voltage dependence of LTP induction becomes steeper at higher stimulation frequencies – even if the number of pulses is kept constant.

A second prediction concerns the shape of the dendritic voltage time course. Suppose that via dendritic voltage clamp, we artificially impose the postsynaptic voltage to follow a square-wave of amplitude $\Delta u$ and duration T. The plasticity behavior will depend on both $\Delta u$ and T, as shown on **Figure 5**. When the presynaptic neuron spikes in the middle of a square pulse of duration T=5 ms, the amount of LTP induced increases with $\Delta u$. For T=15 ms, plasticity will follow an 'inverted u-shape' as a function of voltage amplitude. If now the presynaptic spike is delivered 10 ms after the end of the square pulse, the synapse undergoes LTD.

## Concluding Remarks

We do not claim that elevated voltage in combination with neurotransmitter release is the direct cause of induction of LTP or LTD. Rather our philosophy is that the voltage time course, if experimentally available, is a very good indicator of whether or not synaptic changes are induced in those synapses that have been presynaptically stimulated. In other words, our model describes the Hebbian condition of joint activity of pre- and postsynaptic neuron in a compact form as the interaction of the glutamate trace left by a presynaptic spike with the time course of the postsynaptic voltage. This philosophy does not exclude that a pharmacological block of later steps in the signaling chain could interrupt the LTP/LTD induction or that a direct experimental manipulation of postsynaptic calcium could induce synaptic plasticity in the absence of presynaptic spike arrival or postsynaptic depolarization. Rather our intuition is that, under physiological conditions, the time course of the voltage in the neighborhood of a stimulated synapse is a reliable indicator of the likelihood of that synapse to undergo plasticity. Our leave-one-out cross-validation results (Table 2) show that this intuition can be transformed into a working model to predict the outcome of future plasticity induction experiments given the voltage trace.

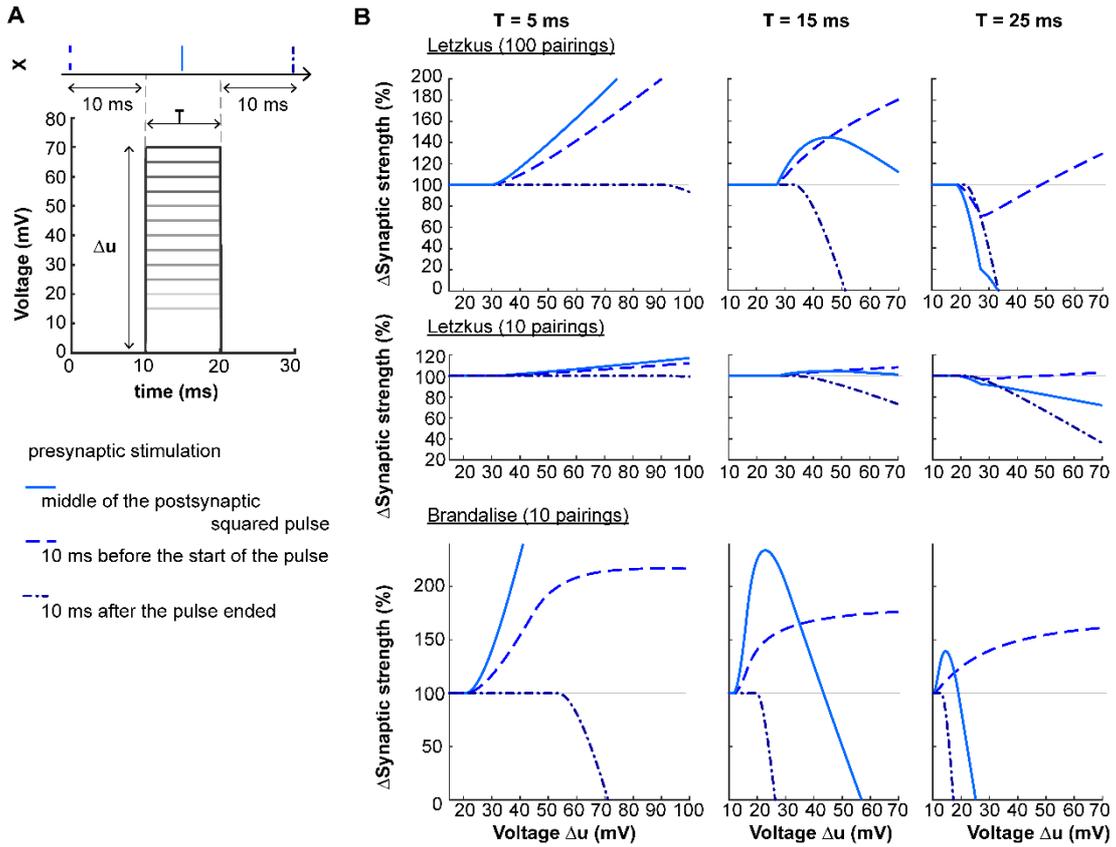

**Figure 5. Non-linear voltage-dependence (A)** The dendritic voltage is clamped for a fixed duration T and varying amplitudes Δu. The resulting squared voltage pulse is paired with a presynaptic spike X arriving 10 ms before the start of the pulse (dashed blue), 10 ms after the end of the pulse (dash-dotted blue), or in the center of the pulse (blue). **(B)** Plasticity as a function of voltage amplitude Δu for T=5 ms (left panels), T=15 ms (middle panels) or T=25 ms (right panels), using two sets of parameters (Letzkus: 100 or 10 pairings, top and middle panels respectively, or Brandalise: 10 pairings, bottom panels, see Table 1).

## 4. Methods

### Voltage-based model of synaptic plasticity

The plasticity model (Figure 1) is a combination of earlier voltage-based models (Brader et al., 2007, Clopath et al., 2010) and the veto concept of Rubin et al. (2005).

Plastic changes of a synapse are caused by potentiation (LTP) or depression (LTD) of the synaptic weight $w$ and add up to a total weight change

$$\frac{d}{dt}w(t) = \frac{d}{dt}w_{LTP} - \frac{d}{dt}w_{LTD}.$$

Potentiation or depression of the weight is induced by a Hebbian combination of presynaptic and postsynaptic activity. Postsynaptic activity is represented by the (low-pass filtered) voltage at the location of the synapse. Presynaptic activity is represented by the spike train $X(t)$ (a sequence of Dirac delta-pulses) arriving at the synapse. The spike train is low-pass filtered and gives rise to a 'trace'

$$\tau_x \frac{d}{dt}\bar{x}(t) = -\bar{x}(t) + X(t)$$

where $\bar{x}$ can be thought of as the amount of neurotransmitter bound to the postsynaptic receptors. The value of $\bar{x}$ increases at the arrival of a spike and decays exponentially with a time constant $\tau_x$ during the interval between spike arrivals (see Figure 1B).

Depression (LTD) is induced if a low-pass filtered version $\bar{u}_-$ of the postsynaptic voltage is above a threshold $\theta_-$ and the "trace of presynaptic activity" $\bar{x}$ does not tend to zero,

$$\frac{d}{dt}w_{LTD}(t) = A_{LTD}\bar{x}(t)[\bar{u}_-(t) - \theta_-(t)]_+$$

where $\bar{u}_-$ is defined as

$$\tau_- \frac{d}{dt}\bar{u}_-(t) = -\bar{u}_-(t) + u(t)$$

with time constant $\tau_-$. The amplitude parameter $A_{LTD}$ characterizes the magnitude of LTD. $[y]_+$ equals $y$ if $y>0$, 0 otherwise.

Potentiation (LTP) is induced if another low-pass filtered version $\bar{u}_+$ of the voltage is above a threshold $\theta_+$ and the "trace of presynaptic activity" $\bar{x}$ does not tend to zero,

$$\frac{d}{dt}w_{LTP}(t) = A_{LTP}\bar{x}(t)[\bar{u}_+(t) - \theta_+]_+$$

where $\bar{u}_+$ is defined as

$$\tau_+ \frac{d}{dt}\bar{u}_+(t) = -\bar{u}_+(t) + u(t)$$

with time constant $\tau_+$. The amplitude parameter $A_{LTP}$ characterizes the magnitude of LTP.

Finally, depression and potentiation compete. If potentiation occurs, the threshold $\theta_-$ increases. The value of $\theta_-$ is determined by the following equation:

$$\theta_-(t) = \theta_0 + \theta(t)$$

with a fixed part $\theta_0$ and a variable part $\theta(t)$ that follows the equation

$$\tau_\theta \frac{d\theta}{dt} = -\theta + b_\theta \frac{d}{dt}w_{LTP}$$

with time constant $\tau_\theta$ and interaction parameter $b_\theta$. This interaction of LTD and LTP parallels the 'veto' concept of Rubin et al. (2005).

We assume that the plasticity framework defined by the above set of equations is generic for glutamatergic NMDA synapses whereas the specific choice of parameters for amplitudes, thresholds and time constants depends on the specific neuron and synapse type as well as on temperature and ion concentrations in the bath of the experimental slice preparation.

## Postsynaptic voltage trace

In the above plasticity model, the value of the postsynaptic voltage at the location of the synapse plays a crucial role. We have access to three experimental datasets where voltage has been measured at a dendritic location close to the synapse (Letzkus et al., 2006; Brandalise et al, 2014 and 2016 and data kindly shared by the authors). Thus, for these plasticity experiments, we do not need to use a neuron model to generate voltage traces; rather, we directly insert a representative experimental voltage trace u(t) into the equations of our plasticity model.

From the Brandalise et al. (2014) dataset, we had access to three cells which had the mean amplitude of their supralinear events, calculated in relation to linear event amplitude increased by at least a factor of two (i.e., difference 100% or more, **Table 5**, column 3). This indicates that the dendritic recording electrode was close enough to the stimulated CA3 recurrent synapses to pick up such a nonlinear effect (see Brandalise et al., 2014). For these cells, the voltage time course combined with plasticity outcome was available for 3 different protocols, which were performed in the following order:

1. no MF stimulation (CA3 alone).
2. the MF and CA3 stimulations occurred at the same time (0 ms)
3. the MF stimulation followed the CA3 stimulation with a 10 ms time interval (+10 ms). The percentage of supralinear events that occurred during this protocol for the 3 individual cells is given in Table 5.

For the remaining protocols (10 ms block, -40 ms and STDP), we used representative voltage time courses and averaged plasticity values, as dendritic recordings and plasticity measurements were done in two different set of cells.

We also model results from Sjöström et al. (2001). In this case, we only had access to representative voltage traces measured at the soma. Since we need for our plasticity model voltage traces in the neighborhood of a synapse, we used the model of L5 basal dendrites from Nevian et al. (2007), available on ModelDB (#124394) to mimic dendritic voltage traces. The multicompartmental model was simulated in NEURON. Action potentials were generated by a 5 ms step current of 3 nA in the somatic compartment and backpropagated through Hodgkin-Huxley-like sodium and potassium channels located on the soma and dendrite. EPSPs were generated by an EPSP-like current injection (double exponential current: 0.5 ms rise time constant, 10 ms decay time constant and peak conductance of 0.1 nS).

The resting potential of all voltage traces (experimental ones and simulation-based ones) has been shifted to 0. This shift allows us to counteract any discrepancies in absolute voltage arising from the electrophysiological recording system or from differences in resting membrane potential across different brain regions and neuron types.

| cell number | rise time (ms) | % increase of amplitude beyond linear | % supralinear events | potentiation (EPSP amplitude change in %) |
|---|---|---|---|---|
| cell 1 | 4.2 | 130 | 34 | 122.0 |
| cell 2 | 3.38 | 100 | 33.3 | 131.0 |
| cell 3 | 6.72 | 140 | 27 | 119.3 |

**Table 5. Characteristics of the +10ms pairing protocol.** Rise time of EPSP, increase in amplitude during supralinear events, percentage of supralinear events and amount of potentiation of the 3 recorded cells. The % increase in amplitude is defined as the difference between the amplitude $a_s$ of the supralinear events and the amplitude $a_l$ of the linear events, divided by the amplitude of the linear events: %= $100(a_s - a_l)/ a_l$. Note that a value of 100 indicates a maximum voltage twice as high as predicted by linear summation.

## Parameter optimization

Our model only defines a mathematical framework whereas specific parameter values may depend on neuron type, synapse type, brain region, as well as details of slice preparations. Therefore, we use different sets of parameters, depending on the experiments we want to model. We take (experimental or simulated) voltage traces as input to our model. Differential equations were solved using forward Euler and with an integration time step of 0.1 ms. Synaptic weights $w$ were initialized at $w_i = 0.5$ and at the end of the simulation we read out the final value $w_f$.

The 9 parameters of our model were fitted to the outcome of different experiments using the Matlab function fmincon (interior-point algorithm). We fixed $\theta_+ > \theta_0$ and defined some upper and lower bounds for the parameters (see **Table 6**). Time constants are in milliseconds with lower bounds always at 2ms and upper bounds below 100ms. In order to mitigate the problem of local minima, we used 25 predefined combinations of parameters as initial points for the optimization algorithm (all inside the bounds). We calculated the least squared error (LSE), which minimizes the quantity SE

$$SE = \sum_{pp} \left( \left[ \frac{w_f - w_i}{w_i} \right]_{pp} - \Delta w_{pp}^{exp} \right)^2$$

where $\Delta w_{pp}^{exp}$ is the experimental plasticity value measured during the protocol $pp$. Since we are interested in the optimal set of parameters, we report in the paper always the parameters from the optimization run which yielded the smallest LSE. We checked that an automatic generation of initial points did not alter the results (Matlab function GlobalSearch).

| bound | $\tau_x$ | $\tau_+$ | $\theta_+$ | $\theta_0$ | $A_{LTP}$ | $A_{LTD}$ | $\tau_-$ | $b_\theta$ | $\tau_\theta$ |
|---|---|---|---|---|---|---|---|---|---|
| lower | 2 | 2 | 8.5 | 2.5 | $10^{-5}$ | $10^{-5}$ | 2 | 0 | 1 |
| upper | 30 | 60 | 30 | 15 | $10^{-2}$ | $10^{-2}$ | 60 | $5.10^5$ | 100 |

**Table 6. Lower and upper bound used during the fmincon search** (same units as in Table 1).

## Data availability

Code is available on GitHub at https://github.com/clairemb90/Voltage-based-model

# 6. Acknowledgements, Author Contributions, Competing Interests statement

**Acknowledgments:** We thank Federico Brandalise and Friedemann Zenke for careful reading and critical comments on the manuscript. We thank Federico Brandalise and Johannes Letzkus for providing additional voltage traces. This research was supported by Swiss National Science Foundation (no.200020_184615) and by the European Union Horizon 2020 Framework Program under grant agreement no. 785907 (HumanBrain Project, SGA2).

This manuscript has been released as a pre-print at https://arxiv.org/abs/2001.03614, Meissner-Bernard et al. (2020).

Author contributions: CMB, WG and LL designed research. CMB performed research and analyzed data. MT contributed unpublished analytic tools. CMB and WG wrote the paper.

Competing Interests statement: The authors declare no competing interest.

# 7. Supplementary Information

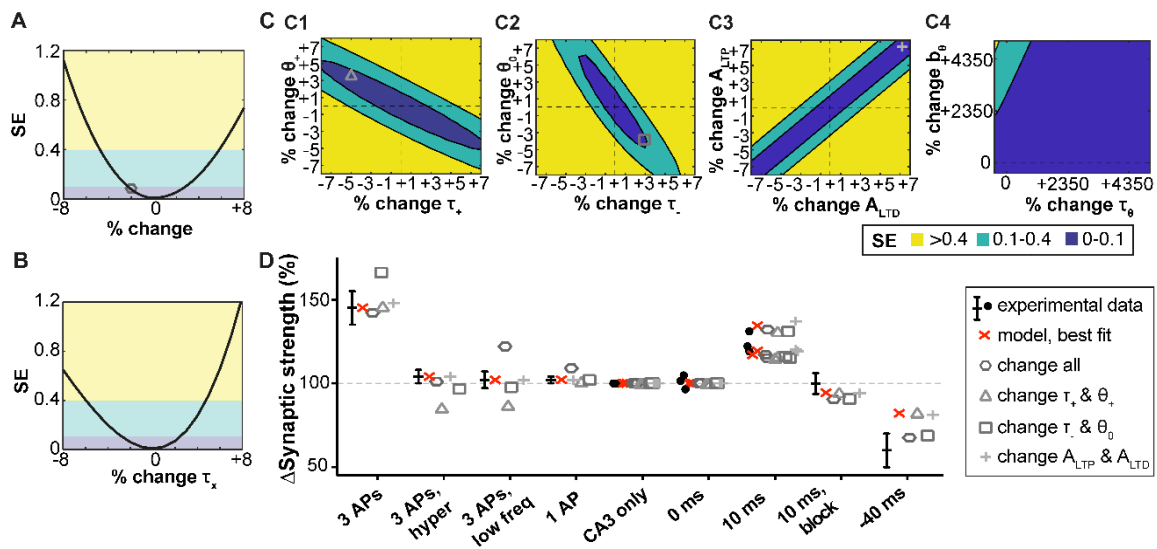

**Supplementary Figure 1. Variation of the squared error (SE) as a function of parameter change.** **(A)** SE (vertical axis) when all the parameters (see Table 1) were increased or decreased (horizontal axis) by a fixed percentage. **(B)** Only the parameter $\tau_x$ is changed (horizontal axis) by a fixed percentage. **(C)** Filled contour plot of the SE while 2 parameters are increased or decreased by a given percentage: $\tau_+$ & $\theta_+$ (C1), $\tau_-$ & $\theta_0$ (C2), $A_{LTP}$ & $A_{LTD}$ (C3) $b_\theta$ & $\tau_\theta$ (C4). **(D)** Plasticity value in 9 different conditions (see Figure 2). Black circles and error bars represent experimental data. Red crosses represent simulations using the parameters obtained with the best fit. Grey symbols in D represent simulations using the parameters obtained with the best fit except a few which were changed by a certain percentage or when all parameters were changed by a fixed percentage (compare symbols in A, C1-C3): hexagon in A (-2%), upwards-pointing triangle in C1 (-5 and +4%), rectangle in C2 (+3 and -4%), cross in C3 (+7 and +7%).